\documentstyle[12pt]{article}
\begin{document}
           \title{\bf Five-dimensional Gravity and the Pioneer
           Effect}
\author{ W. B. Belayev \thanks{dscal@ctinet.ru}\\
\normalsize Center for Relativity and Astrophysics,\\ \normalsize
185 Box , 194358, Sanct-Petersburg, Russia }

\maketitle

\begin{abstract}
In induced gravity theory the solution of the dynamics equations for the test particle on null path leads to additional
force in four-dimensional space-time. We find such force from five-dimensional geodesic line equations and try to
apply this approach to analysis of additional acceleration of Pioneer 10/11, using properties of the asymmetrically
warped space-time.
\end{abstract}

\begin{description}
\item[\it  Keywords:]
induced gravity, brane-world theory, geodesics
\item[\it  PACS:]
04.50.+h, 04.20.Jb, 04.80.Cc
\end{description}

Recently studies \cite{a1,a2,a3,a4}  of five-dimensional
space-time theory analyze consequent departure from 4D geodesic
motion by geometric force. In these papers through 5D null
geodesic motion analysis in induced matter Kaluza-Klein (IM-KK)
gravity theory it was found out appearance extra forces in 4D,
when the scalar potential depends from coordinates of 4D
space-time, and cylinder conditions failed i.e. metric
coefficients depend on the fifth coordinate. Energy-momentum
tensor on 4D sheet is induced through metric dependence on the
extra coordinate. This approach requires 5D vacuum field equations
$\hat{R_{ij}}=0$, where $\hat{R_{ij}}$  is 5D Ricci tensor.
Soliton metric \cite{a5,a6,a7}, which is generalization of the
standard 4D Schwarzschild solution, was obtained in this frame.
More complicated equations are used for description of an
inflationary universe in the Sitter space in Ref. \cite{a8} and a
spherical source with radiation in Ref. \cite{a9}. In brane-world
theory \cite{a10,a11} equations in the brane, contained specific
energy-momentum tensor, should be solved with equations in the
bulk, which are assumed to be 5D Einstein equations with negative
cosmological constant. The brane-world gravity, described by the
bulk metric with warp factor depended from extra coordinate, is
considered, for example, in Refs. \cite{a12} and \cite{a13}.

   In this paper extra forces in 4D space-time, embedded in 5D space-time, are obtained from 5D
null geodesic line equations. We analyze examples of the soliton metric, which describe 5D
space-time with zero curvature, including case of asymmetrically warped space-time, meaning
that the space and time coordinates have different warp factors. We study a possibility of
explanation why the variable component of acceleration, formed the Pioneer effect, exists. We
consider one as consequence of the geometric force (also called fifth force), which is determined
through an explicit metric dependence on extra coordinate of asymmetrically warped space-time.

  Space-time-matter \cite{a1,a2,a7,a11} theory, interpreted fifth coordinate as the rest mass of particles, and
brane \cite{a10,a13,a14} theory have different physical
motivations for the introduction of a large extra dimension.
Though they have same typical scenario and without electromagnetic
potentials one works \cite{a2} for manifold of metrics
\begin{equation}\label{f1}
dS^{2}  = ds^{2} –  \sigma \Phi^{2}(x^{m},y)dy^{2} ,
\end{equation}
where $ds$ is 4D line element, $\Phi$ is scalar potential depended
from 4D coordinates $x^{m}$ and extra dimension $y$, also
$\sigma=±1$. 4D line element is taken in form
\begin{equation}\label{f2}
ds^{2}  = g_{ij}(x^{m},y)dx^{i}dx^{j} ,
\end{equation}
where $g_{ij}$ is metric tensor.

  In IM-KK theory massive particles in 4D have not mass in 5D and move on null path i. e.
$dS=0$. Therefore, 5D particle dynamics equations are found for
null geodesic line just as in 4D by extrimizing \cite{a15}
function
\begin{equation}\label{f3}
 I=\int\limits_{a}^{b}d\lambda\{g_{ij}\frac{dx^{i}}{d\lambda}
 \frac{dx^{j}}{d\lambda}-\sigma \Phi^{2}\frac{dy^{2}}{d\lambda^{2}}\}
 \equiv\int\limits_{a}^{b}hd\lambda  ,
\end{equation}
where $\lambda$ is affine parameter along the path of the particle
terminated at the points $a$, $b$. Null geodesic line equations
are given by
\begin{equation}\label{f4}
\frac{d^{2}X^{A}}{d\lambda^{2}}+\Gamma_{BC}^{A}\frac{dX^{B}}{d\lambda}
 \frac{dX^{C}}{d\lambda}= 0,
\end{equation}
where $X^{A}$ are coordinates of 5D space-time, and
$\Gamma_{BC}^{A}$ are appropriate defined Christoffel symbols. It
should be noticed that if we extremized action with Lagrangian
$L\equiv h^{1/2}$ for particle moving on null path in 5D we would
obtain division by zero, since this movement assigns $h=0$.
Generally choice of parameter $\lambda$ is not arbitrary, and
turned to differentiation with respect to $s$ in Eq. (\ref{f4})we
obtain
\begin{equation}\label{f5}
\frac{d^{2}X^{A}}{ds^{2}}+\Gamma_{BC}^{A}\frac{dX^{B}}{ds}
\frac{dX^{C}}{ds}= -\omega\frac{dX^{A}}{ds},
\end{equation}
where
$\omega=\frac{d^{2}X^{A}}{d\lambda^{2}}/\left(\frac{ds}{d\lambda}\right)^{2}$
and interval $ds$ is assumed to be timelike.

  The first four components of Eq. (\ref{f5}), corresponded to the motion in 4D spacetime, are
transformed to
\begin{equation}\label{f6}
\frac{Du^{i}}{ds}\equiv\frac{d^{2}x^{i}}{ds^{2}}+\Gamma_{jk}^{i}
\frac{dx^{j}}{ds}\frac{dx^{k}}{ds}=f^{i},
\end{equation}
where $f^{i}$ is component of the "extra" force (per unit mass).
Eq. (\ref{f1}) does not set sign of scalar potential and for null
geodesic with spacelike extra coordinate $(\sigma=1)$ yields:
\begin{equation}\label{f7}
 \frac{dy}{ds}=\frac{1}{\Phi}.
\end{equation}
With this condition for metric (\ref{f1}) with 4D line element
(\ref{f2}) fifth force is written as
\begin{equation}\label{f8}
 f^{i}={\mathrm{-}}\frac{g^{ik}}{\Phi}\left(\frac{\partial\Phi}{\partial x^{k}}+
\frac{\partial g_{kj}}{\partial y}u^{j}\right)-\omega u^{i},
\end{equation}
where $u^{j}$ is 4-velocity. The fifth component of Eq. (\ref{f6})
takes the following form:
\begin{equation}\label{f9}
\frac{d^{2}y}{ds^{2}}+\frac{1}{\Phi^{2}}\left(\frac{\partial
g_{ij}}{\partial y}u^{i}u^{j}+2\frac{\partial \Phi}{\partial
x^{i}}u^{i}+\frac{1}{\Phi}\frac{\partial \Phi}{\partial
y}\right)+\omega\frac{dy}{ds}=0
\end{equation}
By substitution (\ref{f7}) in (\ref{f9}) we obtain
\begin{equation}\label{f10}
\omega=-\frac{1}{\Phi}\left(\frac{\partial g_{ij}}{\partial
y}u^{i}u^{j}+\frac{\partial \Phi}{\partial x^{i}}u^{i}\right).
\end{equation}
Then equations (\ref{f8}) are rewritten as
\begin{equation}\label{f11}
 f^{i}={\mathrm{-}}\left(g^{ik}-u^{i}u^{k}\right)\frac{1}{\Phi}\left(\frac{\partial\Phi}{\partial x^{k}}+
\frac{\partial g_{kj}}{\partial y}u^{j}\right).
\end{equation}
When $\Phi=1$, and metric (\ref{f2}) is orthogonal and conforms to
asymmetrically warped space-time:
\begin{equation}\label{12}
ds^{2}= M(y)\tilde{g}_{00}(x^{m})dx^{02}+N(y)\tilde{g}_{ii}(x^{m})
dx^{i2},
\end{equation}
where $M$, $N$ are functions of extra coordinate, components of
fifth force (\ref{f11}) are following:
\begin{eqnarray}\label{f13}
f^{0}=\left(\frac{M'}{M}-\frac{N'}{N}\right)(M\tilde{g}_{00}u^{02}-1)u^{0},
\nonumber \\
f^{i}=\left(\frac{M'}{M}-\frac{N'}{N}\right)M\tilde{g}_{00}u^{02}u^{i},
\end{eqnarray}
where $(')$  denotes derivative with respect to $y$, and in second
equation velocities $u^{i}$ conform to the spacelike coordinates.

  In 5D empty space-time, including 4D sheet with spherical coordinate system $x^{i}=(t,r,\varphi,\theta),$
gravity is described by soliton metric \cite{a7} , whose line
element may be written in form
\begin{equation}\label{f14}
 dS^{2}=H^{a}dt^{2}{\mathrm{-}}H^{-a-b}dr^{2}{\mathrm{-}}H^{1-a-b}d\Omega^{2}{\mathrm{-}}H^{b}dy^{2},
\end{equation}
where $d\Omega^{2}=r^{2}d\varphi^{2}+r^{2}\sin^{2}\varphi
d\theta^{2}$, $H(r)=1{\mathrm{-}}2m/r$, $m$ is mass parameter, and
constants $a$, $b$ satisfy to relation
\begin{equation}\label{f15}
a^{2}+ab+b^{2}=1.
\end{equation}

  Let us analyze modifications of this metric, having parameters conformed to (15), with
coefficients being under constraint of zero Ricci scalar
$\hat{R}.$ That allows to consider one in frame of the induced
matter scenario.
  In first example metric with coefficients, which agree to cylinder conditions, is chosen \cite{a16} in
form
\begin{equation}\label{f16}
 dS^{2}=H^{a}dt^{2}{\mathrm{-}}H^{-a-b}dr^{2}{\mathrm{-}}H^{1-a-b}d\Omega^{2}{\mathrm{-}}H^{b}Qdy^{2},
\end{equation}
where $Q(t)=(1+pt)^{2}$, and $p$ is constant. This space-time has
singularity with $t=–1/p$. In case $a=1$, $b=0$ we have
Schwarzschild limit for metric of embedded 4D space-time. Then
non-vanishing component of the Ricci tensor is following:
\begin{equation}\label{f17}
\hat{R}_{01}=\frac{pm}{r(r-2m)(1+pt)}.
\end{equation}
We note that condition $\hat{R}=0$ is fulfilled with every $a$,
$b$ corresponding equation (\ref{f15}). Taken into account
(\ref{f11}) in considered case Eqs. (\ref{f6}) yield
\begin{equation}\label{f18}
\frac{Du^{0}}{ds}=-\left(\frac{1}{H}-u^{02}\right)\frac{p}{1+pt}
\end{equation}
and
\begin{equation}\label{f19}
\frac{Du^{i}}{ds}=u^{i}u^{0}\frac{p}{1+pt}.
\end{equation}
In these equations  $Du^{j}/ds$ correspond to standard
Schwarzschild metric.

  The next example is metric described asymmetrically warped space-time:
\begin{equation}\label{f20}
 dS^{2}=H^{a}y^{2}dt^{2}{\mathrm{-}}V(H^{-a-b}dr^{2}+H^{1-a-b}d\Omega^{2}){\mathrm{-}}l^{2}H^{b}dy^{2},
\end{equation}
where $V(y)=1+q\ln y$, and $q$, $l$ are constants. Here we have
singularity with $y=e^{-1/q}$ and
\begin{eqnarray}\label{f21}
\hat{R}_{00}=\frac{3qH^{a-b}}{2l^{2}V}, \ \
\hat{R}_{11}={\mathrm{-}}\frac{q^{2}H^{-a-2b}}{4l^{2}y^{2}V}, \ \
\hat{R}_{14}=\frac{m[2(b-a)V+(a+2b)q]}{2(r-2m)ryV}, \nonumber \\
\hat{R}_{22}\sin^{2}\varphi=\hat{R}_{33}={\mathrm{-}}\frac{q^{2}r^{2}H^{1-a-2b}}{4l^{2}y^{2}V}\sin^{2}\varphi,
\ \
 \hat{R}_{44}=\frac{3q(q+2V)}{4y^{2}V^{2}}.
\end{eqnarray}
Expression for the geometric force acting on massive particles can be written as
\begin{eqnarray}\label{f22}
f^{0}=l^{-1}H^{-b/2}\left[\left(\frac{2}{y}-\frac{q}{yV}\right)(y^{2}H^{a}u^{02}-1)+
\frac{blm}{r^{2}}H^{b/2-1}u^{1}\right]u^{0},\nonumber \\
f^{1}=l^{-1}H^{-b/2}\left[\left(2-\frac{q}{V}\right)yH^{a}u^{02}u^{1}+
\frac{blm}{r^{2}}H^{b/2-1}\left(\frac{H^{a+b}}{V}+u^{12}\right)\right],
\nonumber \\
f^{2,3}=l^{-1}H^{-b/2}\left[\left(2-\frac{q}{V}\right)yH^{a}u^{02}+
\frac{blm}{r^{2}}H^{b/2-1}u^{1}\right]u^{2,3}.
\end{eqnarray}
Solution of Eq. (\ref{f7}) in case of invariable $r$, $\varphi$,
$\theta$ gives
\begin{equation}\label{f23}
y=K\exp\left(\frac{t}{l}H^{(a-b)/2}\right),
\end{equation}
where $K$ is constant. Assuming that $y=1$ corresponds to the
background with infinite $r$ and time $t_{0},$ we obtain
\begin{equation}\label{f24}
K=\exp(-t_{0}/l).
\end{equation}
Then the proper time is brought in form
\begin{equation}\label{f25}
\tau=l\exp\left(\frac{t-t_{0}}{l}H^{a-b/2}\right)-
l\exp\left(\frac{-t_{0}}{l}H^{a-b/2}\right).
\end{equation}

  For finding of suggestion on physical interpretation of this approach we apply these results to
analysis of the Pioneer effect. One consists in additional
acceleration of the spacecrafts\cite{a17,a18,a19} towards the Sun.
It follows from Ref. \cite{a18} that force given the Pioneer
effect has two components, namely, constant in considerable area
of the solar system and sinusoid faded with increase distance.
Possible explanation of existence of the constant component is in
Ref. \cite{a19}. In Earth's center coordinate system a velocity,
related to the spacecraft, approximately expressed as
\begin{equation}\label{f26}
u^{1}=u_{\mathrm{p}}^{1}+u_{\mathrm{E}}^{1}\cos(\nu t+\phi_{0}),
\end{equation}
where $u_{\mathrm{p}}^{1}\approx 4.13\cdot 10^{-5}$ conforms to
the radial velocity of the spacecraft \cite{a19} in relation to
the Sun, $u_{\mathrm{E}}^{1}=0.997\cdot 10^{-4}$ conforms to the
orbital velocity of the Earth, $\nu=2\pi$ $\mathrm{y}^{-1}$ is
frequency of the Earth rotation around the Sun, and $\phi_{0}$ is
constant. The best fit of the amplitude of the spacecraft
acceleration periodical component $\tilde{a}_{\mathrm{p}}$
\cite{a18} by means of function $L/r$ gives $L=1.28\cdot 10^{-4} \
\mathrm{sm}^{2}/\mathrm{s}^{2}$.

  When parameters of metric (\ref{f20}) are $a=1$, $b=0$, substitution of fifth force (\ref{f22}) in appropriate
equations (\ref{f6}) with extra coordinate determined by
(\ref{f23}) and (\ref{f24}) gives
\begin{eqnarray}\label{f27}
\frac{du^{1}}{ds}+\frac{mHy^{2}}{r^{2}V}u^{02}-\frac{mH^{-1}}{r^{2}}u^{12}+
(2m-r)(u^{22}+\sin^{2}\varphi u^{32})= \nonumber \\
\left[\frac{(2-q)l+2(t-t_{0})+2tm/r}{(1+q\ln y)l^{2}}\right]
yHu^{02}u^{1}
\end{eqnarray}
for the first component and
\begin{eqnarray}\label{f28}
\frac{Du^{0}}{ds}=l^{-1}\left[\left(\frac{2}{y}-
\frac{q}{yV}\right)(y^{2}Hu^{02}-1)\right]u^{0},\nonumber \\
\frac{Du^{2,3}}{ds}=l^{-1}\left[\left(2-\frac{q}{V}\right)yHu^{02}\right]u^{2,3}
\end{eqnarray}
for the other components with  $Du^{i}/ds$ corresponded to
standard Schwarzschild metric. Let us determine constraints on the
parameters of Eq. (\ref{f27}) in case of radial movement. Assuming
that in the solar system area
\begin{equation}\label{f29}
|(2{\mathrm{-}}q)l+2c(t{\mathrm{-}}t_{0})|\ll\frac{2tGm}{cr},
\end{equation}
where $t_{0}$ is time of the signal receiving, $t$ is time of the
signal sending, $m$ is Sun's mass, $c$, $G$ are light velocity and
gravitational constant, we obtain $t/l^{2}\approx1.6\cdot
10^{-27}\ \mathrm{s/sm}^{2}$ in this area. Hence proposed
condition also requires $t_{0}\gg 4\cdot 10^{15} \ \mathrm{s}$ and
$|l|\gg 1.6\cdot 10^{21}\ \mathrm{sm}$. Then constant $q$ will be
close to $2$ and we must take into account singularity when
$y\approx e^{-1/2}.$

\begin{itemize}
  \item
The author wishes to thank S.N. Manida and A. I. Cigan for
interesting discussions.
\end{itemize}

\end{document}